**Unsupervised cross domain learning with applications to 7 layer segmentation of OCTs**

Yue Wu[1], Abraham Olvera Barrios[2,3], Ryan Yanagihara[1], Irene Leung[2], Marian Blazes[1], Adnan Tufail[2,3], Aaron Lee[1]

[1] University of Washington, Department of Ophthalmology
[2] Moorfields Eye Hospital
[3] University College London

**Introduction**

Deep learning networks have transformed medical image analysis, with many potential clinical applications.[1] Deep learning models can learn representational features to accurately segment anatomic features, predict clinical outcomes, and even suggest treatment approaches.[2,3] Because these models require large amounts of training data, various methods have been developed to address the lack of expert labeled data.[4] Transfer learning, in which the network is pre trained on large general image database and then fine-tuned on a smaller more task-specific dataset, is a useful approach.[5]

One major remaining challenge, however, is the problem of domain shift, where deep learning models trained on a particular dataset show decreased performance when applied to similar but different data. This situation occurs often in the clinical setting, such as the case of images obtained at two different hospitals, under different protocols (dilated vs undilated eyes, patient positioning for chest x-rays), or with a different camera. To date, most strategies to address this problem require additional labeled data from the new setting, which can be challenging, time-consuming, and expensive to obtain[6]. In an ophthalmology clinic, for example, a useful deep learning model should be able analyze optical coherence tomography (OCT) images from different OCT devices, or even older and newer versions of the same device. The first artificial intelligence-based device to obtain FDA approval, which screens patients for signs of diabetic retinopathy based on retinal fundus imaging, illustrates the significance of this problem: the device is only approved for use with one particular camera because of concerns about performance degradation due to domain shift (in this case, images from another camera model).[6,7]

Several methods have been proposed to address this issue. Domain adaptation is a technique in which the new unlabeled data is transformed "in the style" of the training data while preserving the structural information, so that the transformed data is recognizable to the trained model. Generative adversarial networks (GANs) are used to generate synthetic images from the new unlabeled data that resemble the original training data. This approach has been demonstrated using chest x-ray images, by first transforming images from the new data to generate source-like images, and then inputted those into the original segmentation model.8 Here we expand upon that approach by combining these steps into one model, in which the

GAN is nested into the segmentation network, providing real time feedback and improved generalization performance.

**Method**

We developed a cross-domain unsupervised learning method that is similar in philosophy to Cycada(Hoffman et al. 2018). Our model, as shown in Figure 1, combines a GAN, specifically a modified version of U-GAT-IT(Kim et al. 2019), with a supervised model. The U-GAT-IT model was modified to work on 256*256 input images and had the identity or colour channel loss reduced in importance, as our input images were gray scale. For the segmentation results in this paper, the supervised model is a U-Net(Ronneberger et al. 2015) with added dropout layers. Our model was trained end-to-end using the combined U-GAT-IT loss and the supervised loss:

$$L_{total} = L_{UGATIT} + \lambda_s L_{supervised}$$
$$L_{UGATIT} = \lambda_1 L_{lsgan} + \lambda_2 L_{cycle} + \lambda_3 L_{identity} + \lambda_4 L_{cam}$$
$$L_{supervised} = NFL(A, A_{label}) + NFL(A2B, A_{label}) + NFL(A2B2A, A_{label})$$

The supervised loss consists of three terms, and evaluates the combined predictive accuracy of the source domain A, A in the style of the target domain B (A2B), and the reconstructed A from transforming into B and back (A2B2A). In our experiments, we demonstrate the benefits of using a normalized focal loss(Ma et al. 2020) versus the traditional cross entropy loss and also consider simplifying the supervised loss to only the losses on A and A2B. We will refer to the model using a cross entropy loss as GANSL_CE, the model with the normalized focal loss as GANSL_NFL, and the model with the additional penalty on A2B2A as GANSL_NFL_A2B2A.

**Data**

We obtained 110 manually segmented Heidelberg OCT B-scans from ten patients from Chiu et al (Chiu et al. 2015) for our source domain A. The segmentations were for retinal 7 layers: ILM+RNFL, GCL+IPL, INL, OPL, ONL-EZ, EZ-IZ, and RPE. These labeled data were split 80-10-10 at the patient level into training, validation and test. The Heidelberg B-scans had dimensions 496*768 px. These were centered and cropped into 3 non-overlapping images of 256*256 px. This yielded a total of 176 training images, 22 validation and 22 test images for the source domain.

For our target domain, we extracted Topcon 1000 images from the UK Biobank. The Topcon 1000 OCT volumes consist of 128 B-scans each with dimension 650*512 px, covering an biological area of 6mm * 6mm. Again we center cropped the Topcon 1000 B-scans yielding 2 non-overlapping images of 256*256 px. 556 Topcon 1000 B-scans were sampled across the 128 B-scan positions yielding 1112 training images for the target domain.

To validate the accuracy of the unsupervised segmenter, an external test dataset was created by having three human experts manually segment 30 images. This segmentation was done using itk-SNAP with the human experts all trained using the same protocol manual.

**Results**

The inter-observer intersection over union (IOU) metrics for the graders 1, 2 and 3 are shown in Table 1.

The deep learning IOUs to grader 1 are shown in Table 2.

Our deep learning model outperforms existing cross domain segmentation models such as Cycada(Hoffman et al. 2018) as well as more cutting edge models such as MaxSquaresLoss(Chen et al. 2019) in Table 3.

**Discussion**

Our unsupervised cross domain segmentation algorithm achieved comparable IOU versus the human graders. This result means that we can transfer training data across domains, and not be limited to manufacturer or even camera versions, thereby greatly increasing the generalizability of deep learning algorithms for supervised tasks such as classification and segmentation.

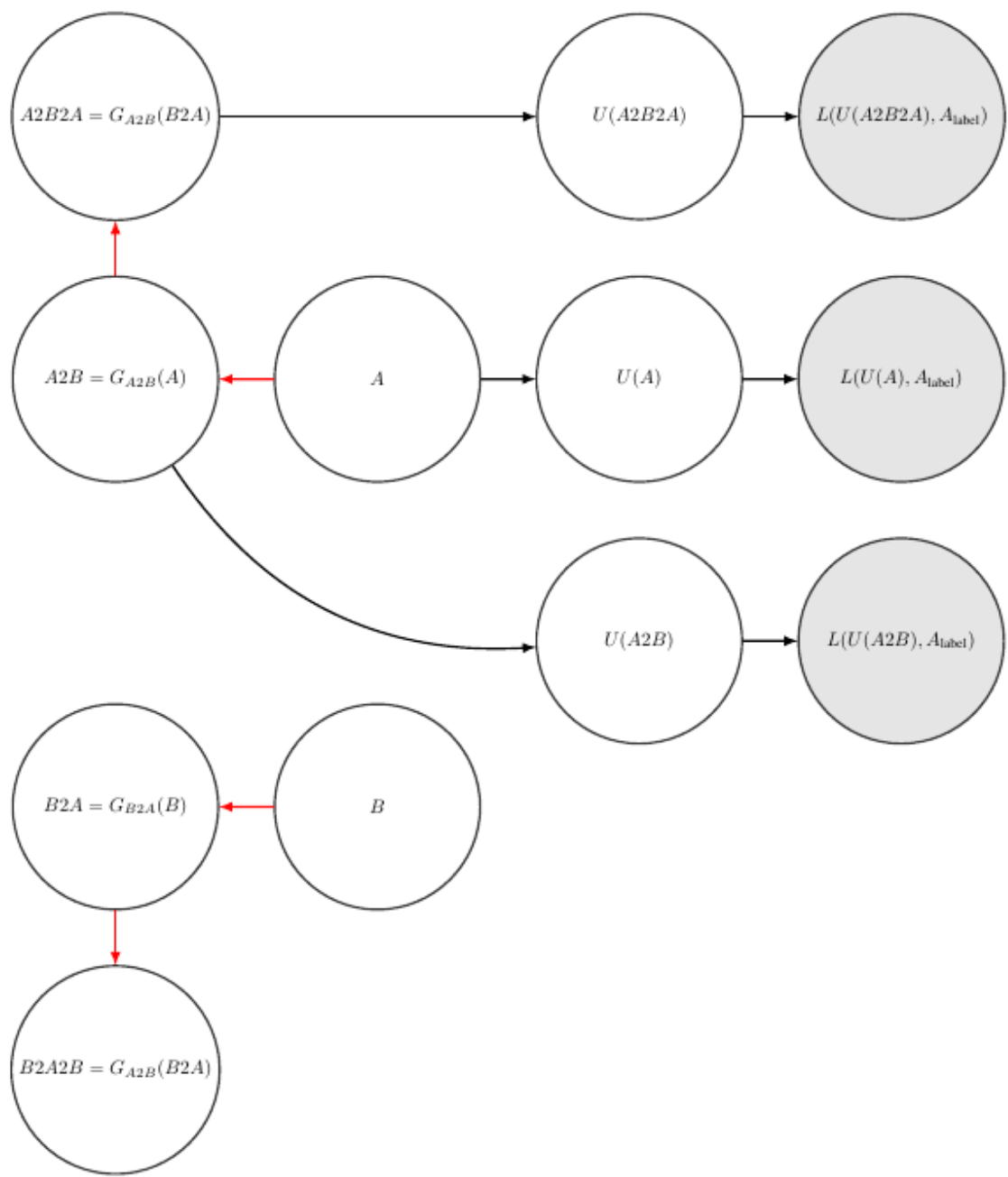

**Figure 1.** Schematic of Deep Learning Architecture

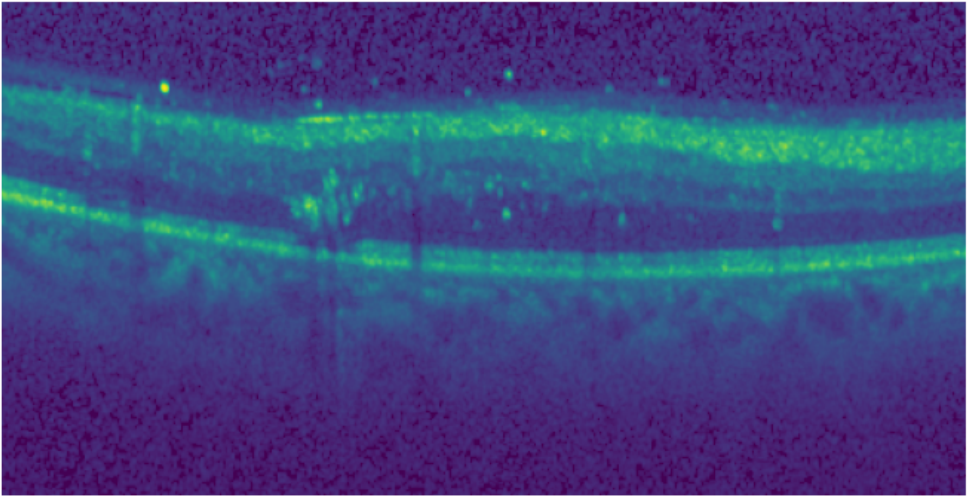

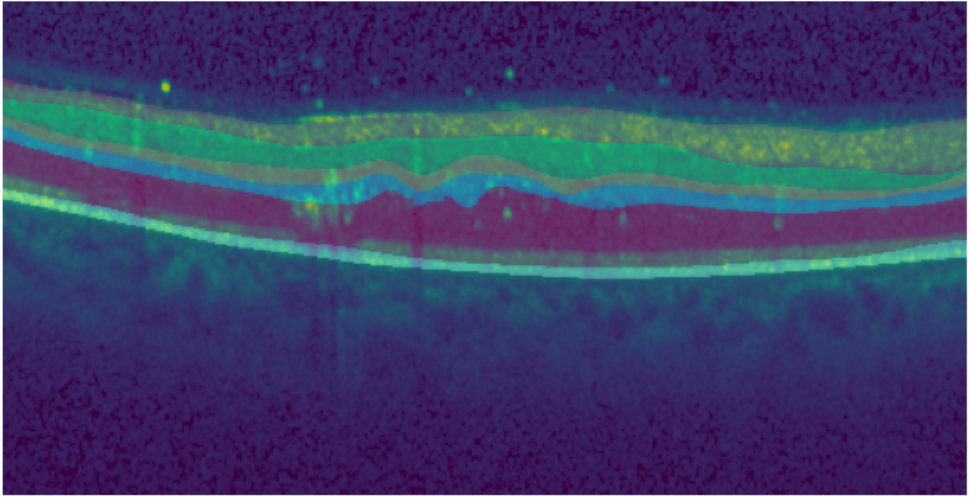

**Figure 2.** A sample Duke OCT B-scan and its corresponding segmented layers

| Layer | G1 vs G2 | G1 vs G3 | G2 vs G3 |
|---|---|---|---|
| BG | 99.7% | 99.5% | 99.6% |
| ILM + RNFL | 80.4% | 75.9% | 72.1% |
| GCL + IPL | 87.7% | 82.2% | 80.6% |
| INL | 71.7% | 58.9% | 57.1% |
| OPL | 65.7% | 62.2% | 59.3% |
| ONL - EZ | 87.4% | 84.8% | 87.9% |
| EZ - IZ | 77.8% | 76.7% | 80.1% |
| RPE | 62.8% | 52.8% | 54.3% |

**Table 1.** Interobserver IOUs for 30 Topcon 1000 images by segmentation layer

| Layer | U-Net_CE | U-Net_NFL | GANSL_CE | GANSL_NFL | GANSL_NFL_A2B2A |
|---|---|---|---|---|---|
| BG | 72.7% | 71.2% | 78.8% | 96.2% | 95.9% |
| ILM + RNFL | 36.5% | 35.4% | 41.7% | 59.8% | 61.0% |
| GCL + IPL | 15.2% | 18.2% | 57.2% | 68.0% | 70.6% |
| INL | 9.9% | 8.4% | 30.6% | 54.9% | 62.0% |
| OPL | 2.8% | 10.3% | 27.9% | 58.5% | 59.3% |
| ONL - EZ | 13.7% | 12.8% | 38.3% | 69.4% | 78.6% |
| EZ - IZ | 14.6% | 0.8% | 20.7% | 25.4% | 27.2% |
| RPE | 14.4% | 0.4% | 33.0% | 47.7% | 60.9% |

**Table 2.** IOU of various deep learning methods to Grader 1

| Layer | Cycada | MaxSquares Loss | GANSL_NFL_A2B2A |
|---|---|---|---|
| BG | 84.0% | 92.4% | 95.9% |
| ILM + RNFL | 19.3% | 46.3% | 61.0% |
| GCL + IPL | 25.6% | 41.1% | 70.6% |
| INL | 16.1% | 41.0% | 62.0% |
| OPL | 17.8% | 34.6% | 59.3% |
| ONL - EZ | 35.5% | 52.2% | 78.6% |
| EZ - IZ | 5.3% | 9.7% | 27.2% |
| RPE | 5.1% | 11.9% | 60.9% |

**Table 3.** Our model vs existing cross domain segmentation models.